\documentclass[letterpaper, 10 pt, conference]{ieeeconf}  % Comment this line out
\IEEEoverridecommandlockouts                              % This command is only
\usepackage{multicol} % This is so we can have multiple columns of text side-by-side

\usepackage[svgnames]{xcolor} % Specify colors by their 'svgnames', for a full list of all colors available see here: http://www.latextemplates.com/svgnames-colors
\usepackage{algorithm}
\usepackage{algpseudocode}
\usepackage{times} % Use the times font
\algnewcommand\algorithmicforeach{\textbf{for each}}
\algdef{S}[FOR]{ForEach}[1]{\algorithmicforeach\ #1\ \algorithmicdo}

\usepackage{graphicx} % Required for including images
\graphicspath{{figures}} % Location of the graphics files
\usepackage{booktabs} % Top and bottom rules for table

\usepackage{amsfonts, amsmath, amsthm, amssymb} % For math fonts, symbols and environments
\usepackage{wrapfig} % Allows wrapping text around tables and figures
\usepackage{epstopdf}

\title{\LARGE \bf
Evolutionary Algorithms for Multi-Objective Optimization of Drone Controller Parameters 
}
\author{ \parbox{2 in}{\centering Azin Shamshirgaran*
        \thanks{* Corresponding author}\\
         Department of CSE \\ %Electrical Engineering and Computer Science\\
         University of California, Merced\\
         Merced, CA 95343\\
         {\tt\small ashamshirgaran@ucmerced.edu}}
         \hspace*{ 0.5 in}
         \parbox{2 in}{ \centering Hamed Javidi\\
         Department of EECS \\ %Electrical Engineering and Computer Science\\
         Cleveland State University\\
         Cleveland, OH 44115\\
         {\tt\small h.javidimostafapourboshroyeh\\@vikes.csuohio.edu}}
         \hspace*{ 0.5 in}
         \parbox{2 in}{ \centering Dan Simon\\
         Department of EECS \\ %Electrical Engineering \\ and Computer Science\\
	Cleveland State University\\
	Cleveland, OH 44115\\
         {\tt\small d.j.simon@csuohio.edu}}
}

\begin{document}

\maketitle
\thispagestyle{empty}
\pagestyle{empty}

\begin{abstract}
Drones are effective for reducing human activity and interactions by performing tasks such as exploring and inspecting new environments, monitoring resources and delivering packages. Drones  need a controller to  maintain  stability  and to  reach  their  goal. The most well-known drone controllers are proportional-integral-derivative (PID) and proportional-derivative (PD) controllers. However, the controller parameters need to be tuned and optimized.
%Drones are effective for minimizing human interactions and can be set up to help and improve the everyday lives of people especially, during the COVID-19 pandemic, or any similar outbreaks in future. 
%\textcolor{blue}{During the current COVID-19 pandemic, and any similar outbreaks in the future, drones can be arranged to help and improve the everyday lives of people. Drones are indeed effective at minimizing human interactions, which is crucial in the time of pandemic. Drones can be used in that direction in two main ways: to facilitate communication between healthcare professionals and persons such as measure their bodies' temperatures and transmit blood or urine samples and also deliver medical supplies to them like medicine or any other healthcare devices.  Drones  need a controller to  maintain  its  stability  in  order  to  reach  their  defined  goal. The most well known controller have been vastly used to control the drones are Proportional-Integral-Derivative ( PID) and Proportional-Derivative ( PD) controller. Also, the controller parameters need to be tuned and optimized.}
%While the goal of single-objective optimization method is finding the minimum value of cost function,
%Also, there are should be an multi objective optimization method to tune controller parameters of drones
%Multi-objective optimization methods is considered in order to tune parameters of a drone's controller such that evaluates several performance metrics of each states of the system such as settling time, overshoot, rise time and steady state error. 
In this paper, we introduce the use of two evolutionary algorithms, biogeography-based optimization~(BBO) and particle swarm optimization (PSO), for multi-objective optimization (MOO) to tune the parameters of the PD controller of a drone. The combination of MOO, BBO, and PSO results in various methods for optimization: vector evaluated BBO and PSO, denoted as VEBBO and VEPSO; and non-dominated sorting BBO and PSO, denoted as NSBBO and NSPSO. The multi-objective cost function is based on tracking errors for the four states of the system.
 %The multi objective function is defined based on the four states of the system which guarantee four important criteria: settling time, overshoot, rise time and steady state error. %The combination of BBO or PSO as an examples of evolutionary algorithms and multi objective optimization ( MOO) results in different algorithms such as 
%This research intends to deeply analyze and compare the results of optimizing this multi-objective problem using vector evaluated and nondominated sorting for both BBO and PSO algorithm. 
Two criteria for evaluating the Pareto fronts of the  optimization methods, normalized hypervolume and relative coverage, are used to compare performance. Results show that NSBBO generally performs better than the other methods.
%Besides the simulation results which show the performance of bio-geography-based optimization comparing to particle swarm optimization algorithm, we investigate the robustness, sensitivity and repeatability of each algorithms regarding to their parameters in detail.

\end{abstract}
\section{INTRODUCTION}
Unmanned aerial vehicles (UAVs), such as drones and quadrotors, have gained significant attention during the last decade.  UAVs can be used in many fields, for instance, inspecting and exploring new environments, monitoring weather patterns to predict tsunamis and earthquakes, construction, monitoring gas and oil resources, and performing jobs in dangerous environments with the advantages of high robustness, reliability, stability, and low resource consumption~\cite{shamshirgaran2016dynamic}, \cite{azin2020}, \cite{cabreira2019survey}.

During the COVID-19 pandemic and similar outbreaks in the future, drones can be set up to improve the everyday lives of people. Drones are effective at reducing human interaction, which is crucial in times of pandemic. To reduce the risk of coronavirus infection, governments have asked and encouraged people to remain in their homes. But then, there should be a way to provide services and support for people in their homes.
Drones can be used for that purpose by facilitating contact-free interactions with healthcare professionals, such as transporting blood or urine samples, and delivering medical supplies like medicine or healthcare devices. During a pandemic, hospitals are potential vectors of contamination, so drones provide an efficient contact-free way to transport critical and necessary medical supplies.  
Although medical supply delivery has been achieved by the commercial company DJI~\cite{djitest}, there are still many challenges, some of which we focus on in this paper. 

 Considering system and environmental noise, many research studies have focused on developing a drone controller to maintain stability, to reach the defined objective, or to tune the controller's parameters. The most well-known controllers for drones are proportional-integral-derivative (PID) and proportional-derivative (PD). Since they are widely applied for drone control, there are a lot of studies on tuning the parameters of these controllers. PID control includes three adjustable gain parameters:
the proportional gain $K_p$, integral gain $K_i$ and derivative gain $K_d$. Algorithms proposed for this tuning problem mostly used aggregation-based multi-objective optimization, which uses a weighted sum of different cost functions to tune the controller parameters. These tuning algorithms require the arbitrary determination of weight coefficients, which may result in an undesirable solution, and they also require high computational effort \cite{chiandussi2012comparison}.

%The proposed algorithm used aggregation multi objective optimization method to tune controller's parameters.  The main advantages of aggregation method are simplicity and computationally efficiency. But it has main disadvantages; for example determining the appropriate weight coefficients is difficult, time-consuming and expensive. Also, it can not guarantee the efficient solution, due to scaling up the cost functions.

Ant colony optimization (ACO), invasive weed optimization (IWO), genetic algorithms (GA), neural networks (NN), particle swarm optimization (PSO) and biogeography-based optimization (BBO) are examples of bio-inspired optimization methodologies that have been used for tuning the parameters of drone controllers~\cite{cespedes2016comparison,khalili2020optimal}. 
%and completely compared for the problem of tuning the parameters of a PID industrial controller for second order plus time delay plants in \cite{cespedes2016comparison}.
In~\cite{boubertakh2009tuning}, ACO was used to tune fuzzy PID controller parameters. The results show performance comparable to PID control, but the proposed method simplified parameter tuning.
In \cite{vishal2014ga}, GA was applied for tuning the PID controller parameters for pitch
control of an aircraft.  
In~\cite{fang2010application, Hamed2014}, an NN was used to tune the PID controller parameters for ship roll reduction. 

We choose BBO and PSO as two algorithms that are representative of a typical evolutionary algorithm and a swarm algorithm, respectively, so we focus on extending these two algorithms to MOO for drone control optimization in this paper. PSO is based on candidate solutions sharing positions in solution space with each other. Each candidate solution, or particle, evolves its position in solution space based on the locations of other particles, until a desirable solution is found~\cite{Cai2007}. BBO is based on islands sharing (or migrating) suitable features, which represent independent variables in the problem solution \cite{simon2008biogeography}. Each island is considered as a possible solution for the problem. Islands gradually evolve by migrating other islands' features to become better habitats (i.e., better solutions) until a desirable solution is found. Both PSO and BBO are able to be implemented as a multi-objective optimization method for a wide variety of applications \cite{javidi2021multi}.
 
Bio-inspired algorithms have been used in previous research to tune drone control parameters. For instance, in~\cite{mohammed2014design}, PSO, bacterial foraging optimization (BFO) and BF-PSO were used to tune PID drone control parameters for roll, pitch and yaw.
In~\cite{mac2016ar}, multi-objective PSO (MOPSO) with an accelerated update methodology was studied to tune PID parameters for the Ar.Drone. They proposed multi-objective functions for the problem and modified the PSO update method for better performance. The objective function was based on settling time $T_s$, rise time $T_r$, overshoot OS and steady state error SSE. They provided experimental results but did not compare their work with other population based algorithms like BBO. Also, they considered an aggregation-based multi-objective cost function, which defines an aggregate cost function as an arbitrarily weighted sum of individual cost functions. In~\cite{boubertakh2013pso}, four decentralized PID controllers were designed to stabilize quadrotor angles and height. A PSO algorithm was used to tune the parameters of the four controllers. Again, this paper did not compare their results with other methods. In~\cite{abdulridha2017planning}, BBO was studied to tune PID parameters to control a hexapod robot to avoid hitches and to follow a wall. This paper considered only the single objective of distance from a wall as the objective function. 
%In \cite{lozovyy2011biogeography}, BBO is used to tune a proportional-derivative control system for sensor selection problem for aircraft engine health estimation. They compare 
%The sate of the art algorithm to tune PID controller parameters is PSO algorithm.

%All these previous algorithm used aggregation multi objective optimization method to tune controller's parameters. The main advantages of aggregation method are simplicity and computationally efficiency. But it has main disadvantages; for example determining the appropriate weight coefficients is difficult, time-consuming and expensive. Also, it can not guarantee the desired solution, due to scaling up the cost functions with different weight coefficients~\cite{chiandussi2012comparison}.

The contribution of this paper is using multi-objective optimization along with evolutionary and swarm algorithms to tune the parameters of the PD controller of a drone. The combination of MOO with BBO or PSO results in four different algorithms which will be the focus of this paper: VEBBO, NSBBO, VEPSO and NSPSO. The multi-objective function is based on the tracking error of the four states of the system. 
Two evaluation criteria, normalized hypervolume and relative coverage of the Pareto front, are used to compare the performance of the methods.

 %In this paper, we intend to measure the performance of PSO and the latest evolutionary algorithm which is BBO in detail by investigating their robustness, sensitivity and repeatability of each algorithms regarding to their parameters to tune PID controller parameters of the drone. 
 
 In Section~II we introduce the dynamic model of the system. In Section~III we explain the BBO and PSO algorithms. In Section~IV we introduce the MOO methods that we combine with BBO and PSO, which include aggregation, VEBBO, VEPSO, NSBBO and NSPSO. In Section~V we compare these methods using two evaluation criteria: normalized hypervolume and relative coverage.

\section{Dynamic Model of the Drone}
\begin{figure}[t]
      \centering
      %\framebox{\parbox{3in}{}}
      \includegraphics[scale=0.4]{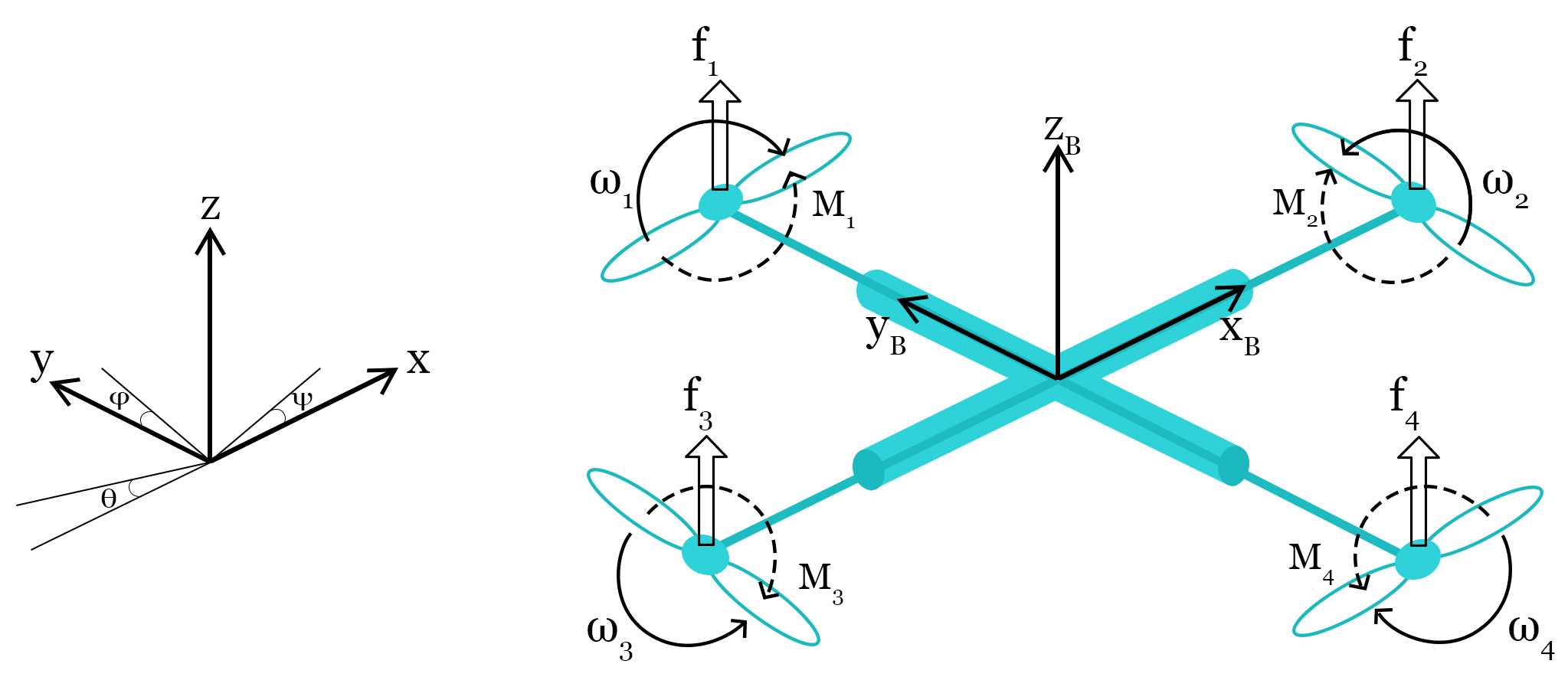}
      \caption{Body frame and inertial frame}
      \label{p1}
   \end{figure}

We used the Euler-Lagrange model to derive the equations of the drone \cite{azin2020}, \cite{bouadi2007nonlinear}, \cite{mellinger2012trajectory}. The linear and angular position of the drone are defined in relation to the inertial reference frame $x$-$y$-$z$ (Figure~\ref{p1}). The angular velocities $p$, $q$, $r$ are defined in relation to the body reference frame $x_B$-$y_B$-$z_B$. The pitch rotation of the drone around the $y$-axis is denoted by $\theta$, the roll rotation around the $x$-axis is denoted by $\phi$, and the yaw rotation around the $z$-axis is denoted by $\psi$. The center of mass of the drone is located at the origin of the body frame. Vector $\epsilon=[x, y, z]^T$ represents linear position, $\eta=[\phi, \theta, \psi]^T$ represents angular position, and $\nu=[p, q, r]^T$ represents angular velocity in the body frame. The drone contains four rotors which induce angular velocities $\omega_i$, torques $M_i$ and forces $f_i$. Thrust $T=f_1+f_2+f_3+f_4$ is created by the combined force in the $z$ axis, and torques $\tau=[\tau_\theta, \tau_\phi, \tau_\psi]^T$ are created in the body frame~\cite{mellinger2012trajectory}, \cite{luukkonen2011modelling}. 
% $I$ defines diagonal Inertia matrix consist of $I_{xx}$, $I_{yy}$, $I_{zz}$. 
 %Figure~\ref{p1} shows the drone structure in body frame and inertial frame~\cite{luukkonen2011modelling},~ \cite{mellinger2012trajectory}.
 Table~\ref{table34} shows the parameters of the drone with their descriptions.

%consisting of $I_{xx}$, $I_{yy}$, $I_{zz}$ 

The linear and angular components of the drone can be defined in two separate subsystems.
The linear components of the system are described as 
%\begin{equation}
 %   \Ddot{\epsilon}=1/m(R_{rot}T-mG
%\end{equation}
\begin{equation}
\begin{bmatrix}
\Ddot{x}\\\Ddot{y}\\\Ddot{z}
\end{bmatrix}
=
\begin{bmatrix}
0\\0\\-g
\end{bmatrix}
+R_{rot}
\begin{bmatrix}
0\\0\\ T/m
\end{bmatrix}
-\frac{1}{m}
\begin{bmatrix}
A_x& 0 &0\\0 &A_y& 0\\0& 0 &A_z
\end{bmatrix}
\begin{bmatrix}
\Dot{x}\\\Dot{y}\\\Dot{z}
\end{bmatrix}
\label{eqn:dy}
\end{equation}
where $A_x$, $A_y$, and $A_z$ denote the drag force coefficients in the $x$, $y$, and $z$ directions respectively of the inertial frame. The rotation matrix $R_{rot}$ is defined in \cite{azin2020}. The angular components of the system are described as 
\begin{equation}
\Ddot{\eta}=J^{-1}(\tau-C(\eta,\Dot{\eta}) \times \Dot{\eta})
\label{eqn:q1}
\end{equation}
where $\tau=[\tau_\theta, \tau_\phi, \tau_\psi]^T$ is the torque in the body frame and $J$ is a positive definite Jacobian matrix; $J(\eta)=W^{T} I W$, where $W$ is a rotation matrix~\cite{azin2020} and $I$ is defined in Table~\ref{table34} as the diagonal inertia matrix containing $I_{xx}$, $I_{yy}$, and $I_{zz}$. The Coriolis matrix $C(\eta,\Dot{\eta})$ is defined in \cite{azin2020}.

\begin{table}[t]

\caption{Drone parameters and their definition}
\label{table34}
\begin{center}
\begin{tabular}{ |c|c|c|c| } 
\hline
Parameters & Definition  \\
\hline
\hline
%\multirow{3}{4em}{Multiple row} & cell2 & cell3 \\ 
%& cell5 & cell6 \\ 
%& cell8 & cell9 \\ 
%\hline
$m$ & Drone mass   \\
\hline
$I$ & Inertia matrix (diagonal) \\
\hline
$l$ & Distance from rotor to center of mass \\
\hline
$\theta$ & Pitch angle; rotation around the y-axis \\
\hline
$\phi$ & Roll angle; rotation around the x-axis \\
\hline
$\psi$ &  Yaw angle; rotation around the z-axis \\
\hline
$O_{x_B y_B z_B}$ & Reference of body frame\\
\hline
$O_{xyz}$ & Reference of inertial frame\\
\hline
$p$, $q$, $r$ & Body frame angular velocities\\
\hline
$x,y,z$ & Inertial frame linear positions \\
\hline
$A_x$, $A_y$, $A_z$ & Coefficients of drag force \\
\hline
$f_i$ & Forces of four rotors \\
\hline
$M_i$ & Torques of four rotors \\
\hline
\end{tabular}
\end{center}
\end{table}

\section{PSO and BBO Algorithms}

\textbf{PSO:} Particle swarm optimization (PSO) is an evolutionary algorithm  based on the concept of collective intelligence in social animals \cite{Cai2007}. The most significant characteristics of PSO are its fast convergence behavior and its inherent adaptability. In PSO, each individual particle of a swarm, which is initially randomly scattered throughout the problem space, can be considered as a potential solution. Particles broadcast their observation, which is based on their current position, to neighboring particles. There are many implementations of PSO. In each iteration of the PSO version that we implement in this paper, there are two important factors that influence particle positions in the next iteration: the personal best position and the global best. The best observation among the previous observations of each particle is called the personal best, and the best previous observation among all particles is called the global best. Individual particles adjust their positions and velocities in proportion to the global best and personal best. 

\begin{equation}
\label{eq:q4}
\begin{split}
  v_{i,d}(t+1) = & wv_{i,d}(t)+ c_1r_1{_{i,d}}(t)(y_{i,d}(t)-x_{i,d}(t))+ \\ & c_2r_{2_{i,d}}(t)(\hat{y}_{d}(t)-x_{i,d}(t))
  \end{split}
\end{equation}

\begin{equation}
\label{eq:q5}
x_{i,d}(t+1)= x_{i,d}(t)+v_{i,d}(t+1) 
\end{equation}

\noindent where $v_{i,d}(t)$ is the velocity of the $d$-th dimension of particle $i$ at iteration $t$, $x_{i,d}(t)$ is its position, $y_{i,d}(t)$ is its personal best position, and $\hat{y}_{d}(t)$ is the best global position of the swarm. Parameters $c_1$, $c_2$, and $w$ are constants (tuning parameters), and $r_1{_{i,d}}(t)$ and $r_2{_{i,d}}(t)$ are random numbers that are uniformly distributed between 0 and 1.

% \begin{algorithm}
%     \caption{PSO Pseudo-code}\label{algor}
%     \begin{algorithmic}[1]
%         \Function{PSO}{$(x_1, x_2,..., x_n),\newline
%         upper-bound_value(x_1, x_2,..., x_n),\newline
%         lower-bound_value(x_1, x_2,..., x_n)$\newline
%         }
%         \State Initialize( population and parameters )
%         \While{($Termination \ criterion \ is \ unsatisfied$)}
%             \For{$i=1 \: to \: Population \: Size$}
%                 \State Calculate particle velocity (see equation \ref{eq:q4})
%                 \State Update particle position (see equation \ref{eq:q5})
%                 \If{$f(X_i) < f(Pb_i)$}
%                     \State $Pbi =  Xi$
%                     \If {$f(Pb_i) < f(Gb)$}
%                         \State $Gb =  Pb_i$
%                     \EndIf
%                 \EndIf
%             \EndFor
%         \EndWhile
%         \EndFunction
%     \end{algorithmic}
% \end{algorithm}

%Algorithm \ref{algor} shows the process of optimizing $x_1, x_2, ..., x_n$ parameters with specific upper bound and lower bound values for each parameters using PSO method. In this algorithm, $f(X_i)$ is a defined objective function that gets $i$th population and $Pb_i$ and $Gb$ are personal best of particle number $i$ and the global best, respectively. 
\textbf{BBO:}
BBO is a recent heuristic algorithm that was first introduced in \cite{simon2008biogeography}. Its performance has proven to be competitive with other optimization algorithms, e.g., differential evolution, GA, PSO, stud GA, and many other algorithms, which motivates us to apply BBO to optimize the PD drone controller and then compare its results with PSO.
BBO works based on sharing suitable features of candidate problem solutions in the way that biological species on islands randomly emigrate and immigrate. Each island is considered as a possible solution to the problem. Islands gradually evolve their features to become better-suited habitats until a desirable habitat is found, which corresponds to a desirable problem solution. 
In BBO, the environment is an archipelago and each island represents a possible solution to the problem.  

% \begin{algorithm}
%     \caption{BBO Pseudo-code}\label{algor2}
%     \begin{algorithmic}[2]
%         \Function{BBO}{$(x_1, x_2,..., x_n),\newline
%         upper-bound_value(x_1, x_2,..., x_n),\newline
%         lower-bound_value(x_1, x_2,..., x_n)$\newline
%         }
%         \State Initialize Population 
%         \State Define Mutation probability  $\mu_{k}$
%         \While{Termination criterion is unsatisfied}
%             \ForEach{$x_{k} \in \ [1,N] $}
%                 \State  Calculate the immigration rate $\mu_{k}$ and emigration rate $\lambda_{k}=1 - \mu_{k},\ \mu_{k} \in [0,1]$ 
%                 \State $\{z_{k}\} \leftarrow \{x_{k}\}$
%             \EndFor
%             \ForEach{ individual $z_{k}$ }
%                 \ForEach{ solution features }
%                     \If{$randnum() < \lambda_{x_{k}}$}
%                         \State Pick a solution from which to emigrate
%                         \State Migrate()
%                     \EndIf
%                     \State update$\ {z_{k}}$
%                 \EndFor
%             \EndFor
%         \EndWhile
%         \EndFunction
%     \end{algorithmic}
% \end{algorithm}

%In the algorithm \ref{algor2}, input parameters are the same as the PSO algorithm. 
%$\mu_{k}$ and $\lambda_{k}$ are the emigration and immigration probability of island $k$. 

\section{Multi-Objective Optimization}
While the goal of single-objective optimization is finding the minimum value of a cost function, multi-objective optimization methods consider several performance metrics, which in our case are captured by the tracking errors of the four states of the system. In general, the solution of a multi-objective optimization problem is a set of points which is known as a Pareto set where no solution is dominated by any other solution. 
The combination of MOO with either BBO or PSO results in different algorithms: vector evaluated BBO (VEBBO), non-dominated sorting BBO (NSBBO), vector evaluated PSO (VEPSO), and non-dominated sorting PSO (NSPSO). VEBBO and VEPSO are based on using one objective function for each recombination to
create the population at the next generation. In NSBBO and NSPSO, the cost function of each individual is
assigned based its domination level. These methods are explained in more detail in \cite{simon2008biogeography}.

The simplest MOO approach is to combine all objective functions into a single scalar objective function, in which case we do not consider Pareto optimization. In Section~\ref{sec-Aggregation} we discuss the aggregation method for drone PD parameter optimization. We then discuss multi-objective Pareto optimization using VEBBO and VEPSO in Section~\ref{sec-VE}, and NSBBO and NSPSO in Section~\ref{sec-NS}.

\subsection{Aggregation Method}
\label{sec-Aggregation}

The main advantages of the aggregation method are simplicity and computational efficiency. But it also has disadvantages; for example, determining the appropriate weight coefficients is difficult and time-consuming. The choice of the weight functions is somewhat arbitrary and may not result in a desirable solution~\cite{chiandussi2012comparison}.
%Both PSO and BBO algorithms which are introduced in the previous sections be able to handle single objective functions. Since the actual objective function of the controller is a multi objective function, in order to apply PSO and BBO to our controller, we need to leverage multiobjective particle swarm optimization (MOPSO) and Multiobjective Biography based optimization (MOBBO) which are able to deal with multiobjective optimization problems.
The aggregated objective function based on the four most important criteria in the PD controller is defined as 
 $$ F(X)= \omega_1 F_1(X) + \omega_2 F_2(X) + \omega_3 F_3(X) + \omega_4 F_4(X) $$ \\
where $w_i$ is the weight for each individual objective function and since we desire the impact of each individual cost function to be the same, $w_i$ is set equal to 1. Each component of $F(X)$ is defined as
\begin{equation}
\label{eq:q8}
\begin{split}
  %F_1(X) = T_r\\
  %F_2(X) = OS\\
  %F_3(X) = T_s\\
    F_{i}(X) = \int_{t}|X_i-X_{i,d}| \, dt
   % F_5(X) = \int_{t}|T|
  \end{split}
\end{equation}
where $i \in [1, 2, 3, 4]$ indexes the objective, the state $X_i \in \{\phi, \theta, \psi, z\}$, and the desired (reference) state $X_{i,d} \in \{\phi_d, \theta_d, \psi_d, z_d\}$. $t$ represents the integration time, which indicates the time of the drone simulation. The objective function is set to minimize the difference between actual and desired value of the state which in turn reduces settling time, overshoot, rise time and steady state error.

\subsection{Vector Evaluated Evolutionary Algorithms}
\label{sec-VE}

One common way to solve multi-objective optimization problems is by keeping a collection of the best solutions in a repository and updating the repository each iteration. In this method, the best solutions are defined as non-dominated solutions or Pareto optimal solutions \cite{coello2000updated}. 

VEBBO is based on using one objective function at a time for recombination to
create the population at the next generation \cite{mohammadi2016multi}.  This method is explained in more detail in~\cite{simon2013evolutionary}.

VEPSO evaluates each candidate solution using only one of the objective functions of the problem. Then, information based on this single objective function is communicated to the other members of the swarm as a representation of its best solution~\cite{parsopoulos2004multiobjective}. 
In VEPSO, several swarms are employed to search the space and information is exchanged among them \cite{Parsopoulos2004RecentAT}. Each swarm is exclusively evaluated with one of the objective functions, but information coming from other swarms is used to influence its motion in the search space. The best position found by each particle separately, as well as the best among these positions, are the main guiding mechanisms of the swarm. Communicating this information between swarms leads to Pareto optimal solutions.

\subsection{Non-Dominated Sorting Evolutionary Algorithms}
\label{sec-NS}

In NSBBO, the cost function of each individual is
assigned based on its domination level. This method is explained in more detail in \cite{simon2013evolutionary}.

NSPSO compares all particles, while considering their positions in the search space both before and after their position updates, in terms of their personal best~\cite{Li2003ANS}. NSPSO leverages non-dominated sorting and two parameter-free niching methods.
At each iteration, NSPSO performs non-dominated sorting for all particles which are distributed in a number of subsets in the main set (population). In the next step, niche count \cite{deb2001multi}, global best, velocity, and position  are computed for each particles in each subset. Then, novel population of size 2N is generated and sorted. Finally, a novel set of N solutions is created by picking fronts particles in each sorted subset.

%\section{Evaluation Criteria}

\section{Simulation Results and Comparisons Between Different MOO methods}
We evaluate the performance of MOBBO and MOPSO on the drone controller via computer simulation using MATLAB/Simulink.
We used the following parameters for MOBBO.\\
\begin{equation}
\label{eq:10}
\begin{split}
  p_s= 50,\; I_t= 30, \; N_e= 2, \;  I_r= 1- E_r\\
  %Emigration-Rate  = (Population-Size + 1 - (1:Population-Size)) \over (Population-Size + 1)\\
  E_r= ( p_s  +1 - f_s) / ( p_s + 1); f_s\in[1,2,...,p_s]
  \end{split}
\end{equation}
where $p_s$, $I_t$, $E_r$, $I_r$ and $N_e$ are population size, iteration limit, emigration rate, immigration rate, and number of elites. $N_e$ is the number of the best solutions to keep from one generation to the next. For MOPSO,
\begin{equation}
\label{eq:11}
\begin{split}
  p_s = 50,\;
  I_t = 30,\;
 % Inertia-Weight = 1\\            % Inertia Weight
w=0.5, \\
w_d= 0.99, \;    % Inertia Weight Damping Ratio
 c_1= 2, \;
c_2= 2
  %Emigration-Rate  = (Population-Size + 1 - (1:Population-Size)) \over (Population-Size + 1)\\
  \end{split}
\end{equation}
where $p_s$, $I_t$, $w$, $w_d$, $c_1$ and $c_2$ are population size, 
iteration limit, inertia weight, inertia weight damping ratio, personal learning coefficient and global learning coefficient.
The desired position $z_d$ of the drone is fixed at~$z_d = 0$, and the desired angular positions are fixed at
$\theta_d = \phi_d = \psi_d = 0$ with the initial  positions and Euler angles selected as $\begin{bmatrix}
x_0, & y_0, & z_0  \end{bmatrix}^T =\begin{bmatrix}
0, & 0, & -1  \end{bmatrix}^T$, and $\begin{bmatrix}
\theta_0, & \phi_0, & \psi_0  \end{bmatrix}^T = \begin{bmatrix}
-0.7, & -0.7, & -0.7  \end{bmatrix}^T$. The drone parameters in our simulations are shown in Table~\ref{table197}.
 \subsection{Simulation Results of Aggregation Method}
\begin{figure}
      \centering
      %\framebox{\parbox{3in}{}}
      \includegraphics[scale=0.17]{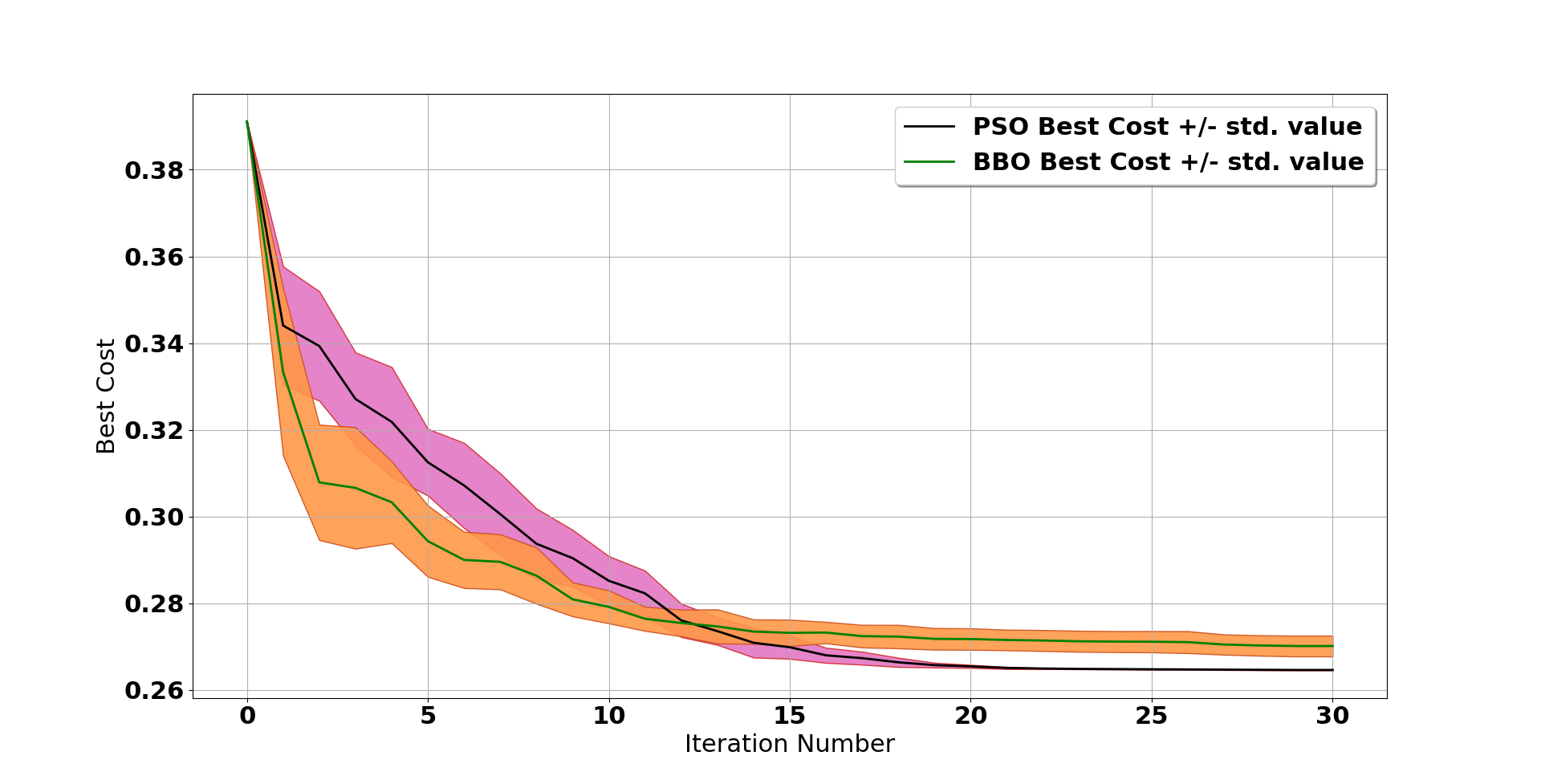}
      \caption{Mean and standard deviation (shaded) of best cost of 5 trials of BBO and PSO}
      \label{p2}
\end{figure}

To simplify notation, we refer to the BBO aggregation method as simply BBO, and the PSO aggregation method as simply PSO. 
Figure~\ref{p2} depicts the mean and standard deviation of the best cost function value for each algorithm over 30 iterations for 5 trials. BBO converges faster, but the mean value of the two algorithms are almost the same at the final iteration; 0.2646 for PSO and 0.2701 for BBO. These results are more than 30\% better (smaller) than the conventional PD controller cost, which is 0.3911, which was obtained using the PD parameters from~\cite{luukkonen2011modelling}, and which provided the initial PD values for both PSO and BBO. In Table~\ref{tabel11}, the Min and Max columns show the search space bounds for the PD parameters, the PD column shows the values used for the conventional PD controller which are chosen manually~\cite{luukkonen2011modelling}, and the PSO and BBO columns show the mean values of the PD parameter that PSO and BBO converged to after 30 iterations and 5 trials. 

\begin{figure}
      \centering
      %\framebox{\parbox{3in}{}}
      \includegraphics[scale=0.17]{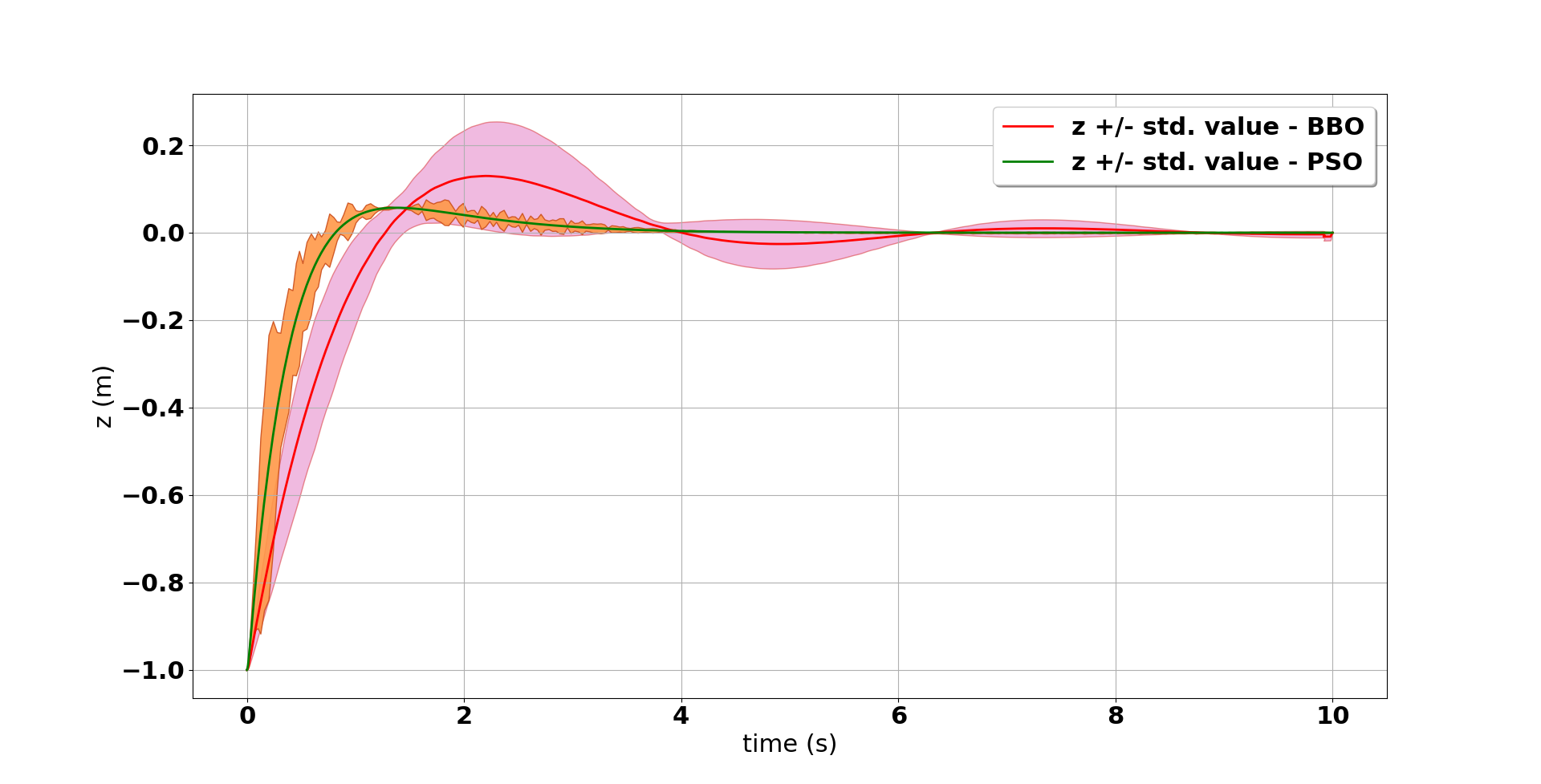}
      \caption{Mean and standard deviation (shaded) of $z$ of 5 trials of BBO and PSO}
      \label{p4}
\end{figure}

Figure~\ref{p4} shows the mean and standard deviation of state $z$ for both PSO and BBO with 0.05 and 0.1 m overshoot, respectively. PSO has a better rise time, about 1 sec, and a better settling time, about 3 sec.

\begin{figure}
      \centering
      \includegraphics[scale=0.17]{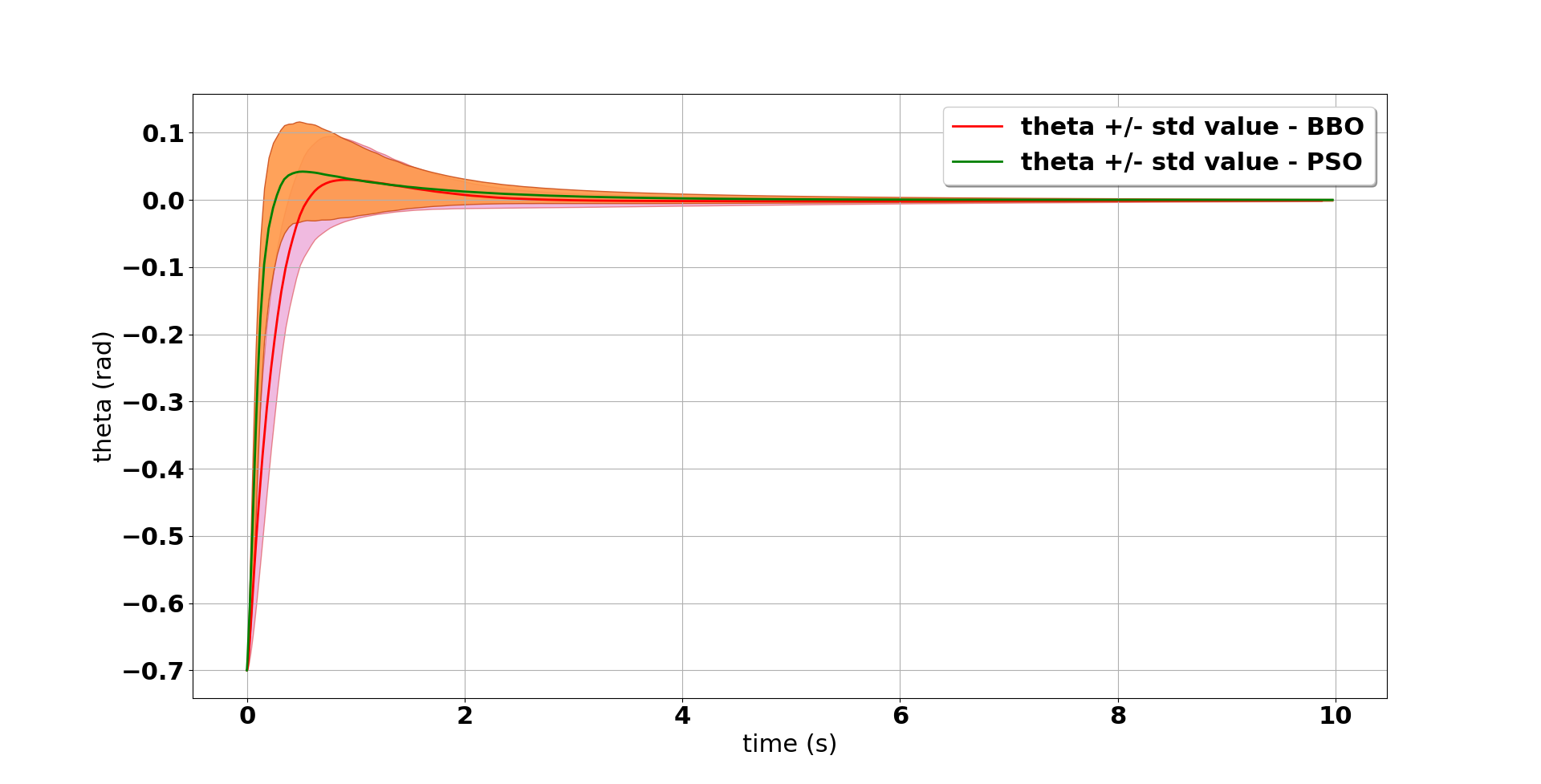}
      \caption{Mean and standard deviation (shaded) of $\theta$ of 5 trials of BBO and PSO}
      \label{p5}
\end{figure}

\begin{figure}
      \centering
      \includegraphics[scale=0.17]{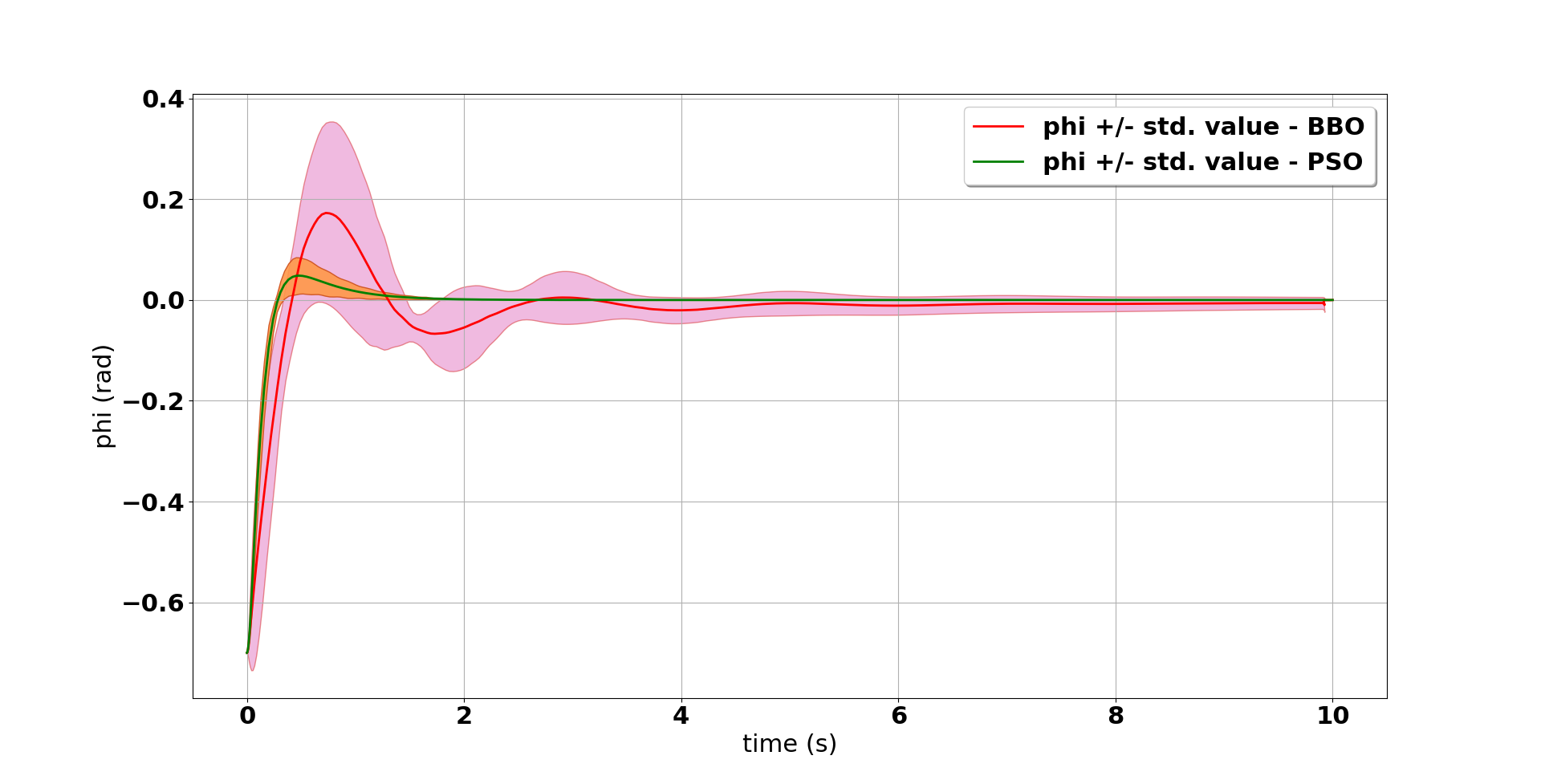}
      \caption{Mean and standard deviation (shaded) of $\phi$ of 5 trials of BBO and PSO}
      \label{p7}
\end{figure}

Figure~\ref{p5} illustrates approximately the same performance for BBO and PSO in terms of $\theta$, showing an overshoot of about 0.05 rad, a rise time of about 0.5 s, and a settling time of about 1.5 s. However, Fig.~\ref{p7} shows that PSO gives better results in terms of both overshoot and settling time for $\phi$: the settling time for BBO is about 4 s while PSO settles in about 2 seconds, and the overshoot for BBO is about 0.2 rad while PSO overshoots about 0.1 rad. 

\begin{figure}
      \centering
      \includegraphics[scale=0.17]{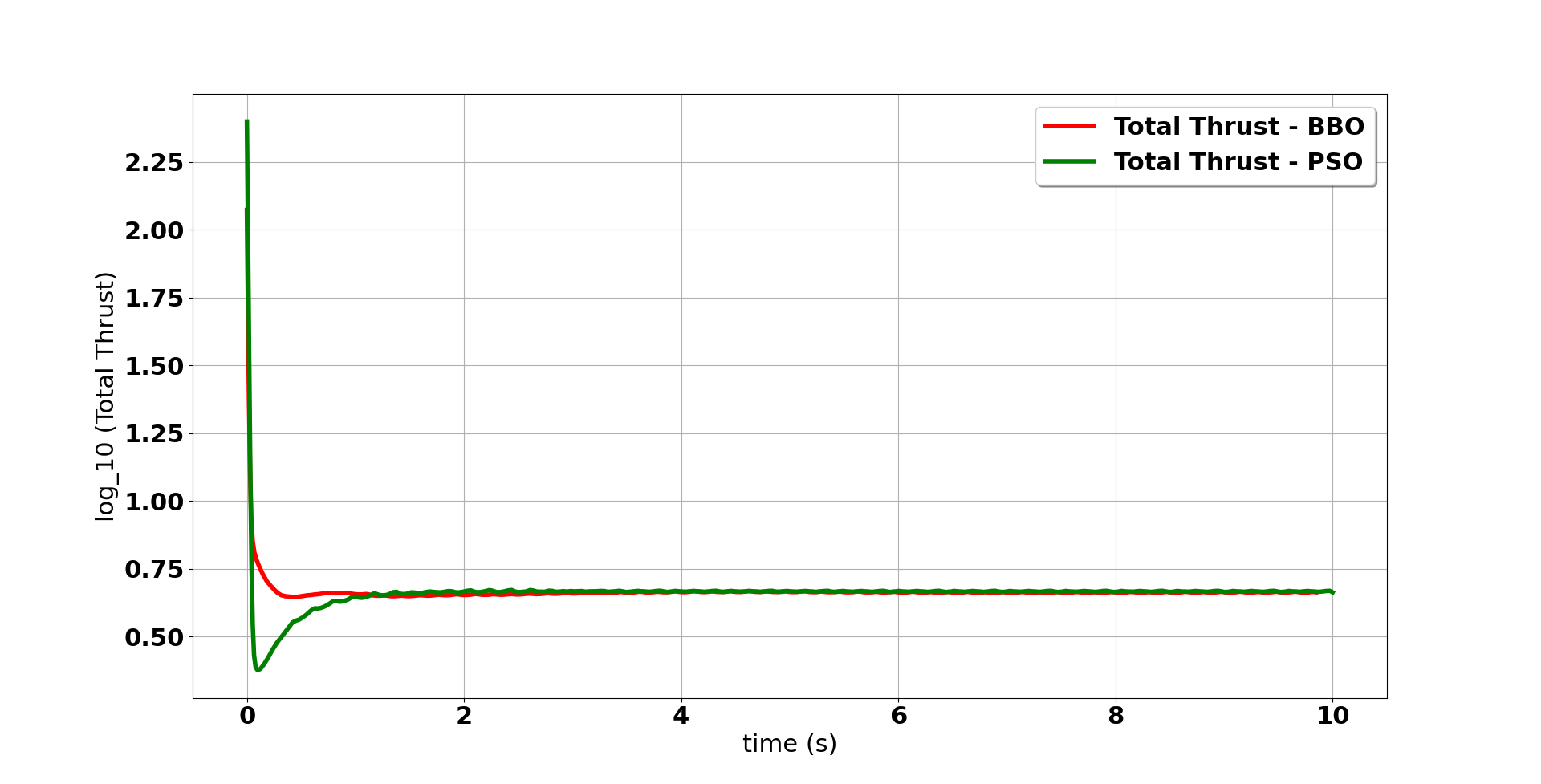}
      \caption{Mean of thrust of 5 trials of BBO and PSO}
      \label{p8}
\end{figure}

Figure \ref{p8} shows the $log_{10}$ of mean of total thrust for 5 trials of BBO and PSO. We used $log_{10}$ to better visualize the differences between PSO and BBO. The beginning of the simulation shows a large spike due to initialization, and trying to move the drone from $z_0=-1$ to $z_d=0$.  

\begin{table}
  \caption{ Parameter values for simulation}
\label{table197}

\begin{center}
\begin{tabular}{ |c|c|c|c| } 
\hline
%\multirow{3}{4em}{Multiple row} & cell2 & cell3 \\ 
%& cell5 & cell6 \\ 
%& cell8 & cell9 \\ 
%\hline
$m$ & 0.468 kg   \\
\hline
$g$ & 9.81 m/s$^2$ \\
\hline
$l$ & 0.225 m \\
\hline
$I_{xx}$ & $4.856 \times 10^{-3}$  kg$\cdot$m$^2$ \\
\hline
$I_{yy}$ & $4.856 \times 10^{-3}$  kg$\cdot$m$^2$ \\
\hline
$I_{zz}$ & $8.801 \times 10^{-3}$  kg$\cdot$m$^2$ \\
\hline
$A_x$ & 0.25 kg/s \\
\hline
$A_y$ & 0.25 kg/s )\\
\hline
$A_z$ & 0.25 kg/s \\
\hline
\end{tabular}
\end{center}
\end{table}

\begin{table}
\caption{PD tuning comparison. The first subscript, $p$ or $d$, indicates the proportional or derivative gain. The second subscript, $\phi$, $\theta$, etc., indicates the state of the system.}
\label{tabel11}
\begin{center}
\begin{tabular}{ |c|c|c|c|c|c| } 
\hline
 & Min & Max & PD & PSO & BBO  \\
\hline
\hline
%\multirow{3}{4em}{Multiple row} & cell2 & cell3 \\ 
%& cell5 & cell6 \\ 
%& cell8 & cell9 \\ 
%\hline
$K_{p_{\phi}}$ &0&20.0&6&14.015&19.7704\\
\hline
%$K_{i_{Phi}}$ &0&0.1&0&0.0015&0.0022\\
%\hline
$K_{d_{\phi}}$ &0&10&1.75&10&9.6322\\
\hline
$K_{p_{\theta}}$ &0&10&6&2.7624&3.04\\
\hline
%$K_{i_{Theta}}$ &0&0.3&0&0.2824&0.2413\\
%\hline
$K_{d_{\theta}}$ &0&10&1.75&10&9.6105\\
\hline
$K_{p_{\psi}}$ &0&10&6&6.4304&1.49\\
\hline
%$K_{i_{Psai}}$ &0&0.3&0.1&0.2921&0.012\\
%\hline
$K_{d_{\psi}}$&0&10&1.75&10&9.9945\\
\hline
$K_{p_{Z}}$&0&3& 1.5 & 3 & 2.8141 \\
\hline
%$K_{i_{Z}}$&0&0.1& 0.1 & 0.1291 &  0.016\\
%\hline
$K_{d_{Z}}$& 0&3&2.5 & 2.7755 & 2.89\\
\hline
\end{tabular}
\end{center}
\end{table}

\subsection{Parallel Computation for Aggregation Method}
To enhance computational efficiency, we use the MATLAB Parallel Computing Toolbox. For each iteration, the BBO algorithm needs to run a simulation for each individual in the population, along with its PD parameters, and then compute the cost. This operation can be done in parallel rather than sequentially to save time. Table~\ref{table97} shows the time savings when we run BBO with and without the parallel computing toolbox. This approach to saving time can be especially helpful in more complicated problems. For this scenario, we use population size $p_s=32$, iteration limit $I_t=30$ and linear emigration. The CPU is an \textit{Intel(R) Core(TM) i5-8300H} with 8 logical processors (4 physical and 4 virtual). The population size should ideally be a multiple of the number of physical cores because each core works independently and runs its own simulation. Here, we have a population size of 32, so each core (worker) handles 8 simulations, i.e., 8 BBO individuals.

\begin{table}
 \caption{Run time for sequential and parallel BBO}
\label{table97}
\begin{center}
\begin{tabular}{ |c|c| } 
\hline
 & Run Time (s)  \\
\hline
\hline
Sequential BBO & 1505 \\
\hline
Parallel BBO & 930 \\
\hline
\end{tabular}
\end{center}
\end{table}
 \subsection{Sensitivity, Robustness and Repeatability for Aggregation Method}
In this section, we provide some statistical results to compare BBO and PSO in detail. 
First, we check the repeatability of PSO and BBO over 10 trials; Table~\ref{table52} shows the minimum, maximum and average objective function value of these 10 trials. The standard deviation (Std Dev) column shows that there are small differences between the results of the 10 trials, which shows small variability or dispersion around the average. 
Next, we check the sensitivity of BBO and PSO to their parameters, like population size. Table~\ref{table51} shows the cost for different population sizes, where the number of iterations remained equal to 30 for each population size. The population size affects BBO more than PSO. 

Another parameter that we investigate is the emigration rate function and its effect on BBO cost. Table~\ref{table57} shows the average cost and stand deviation values for 10 runs of the algorithm for two different kinds of emigration functions. The table shows that the linear emigration function that we introduced in Equation~\ref{eq:10} works better and gives a better cost that the sinusoidal emigration function~\cite{simon2013evolutionary}.

For PSO, we have three different variables for which we check sensitivity: inertia weight $w$, personal learning coefficient $c_1$ and global learning coefficient $c_2$. Finding the appropriate values for $c_1$ and $c_2$ is tricky and we used the recommended baseline values in~\cite{simon2013evolutionary}. Table~\ref{table66} shows PSO performance for different values of $c_1$ and $c_2$ for 10 simulations. This table shows that the cost values are relatively independent of the values of $c_1$ and $c_2$. That is, there is no need to put much effort into finding appropriate values of $c_1$ and $c_2$, as long as they remain within well-established limits. Table~\ref{table45} shows the sensitivity of PSO to inertia weight. This table shows that PSO is not sensitive to inertia weight, as long as it remains within well-established limits~\cite{simon2013evolutionary}.

\begin{table}
  \caption{Repeatability of BBO and PSO over 10 trials}
\label{table52}
\begin{center}
\begin{tabular}{ |c|c|c|c|c| } 
\hline
 & Min & Max & Avg & Std Dev  \\
\hline
\hline
BBO & 0.2674 &  0.2804 & 0.2713  & 0.0040  \\
\hline
PSO & 0.2645 & 0.2672 & 0.2656 & 0.0009 \\
\hline
\end{tabular}
\end{center}
\end{table}

\begin{table}
\caption{Sensitivity of BBO and PSO to population size}
\label{table51}
\begin{center}
\begin{tabular}{ |c|c|c|c| } 
\hline
Population Size & BBO Cost & PSO Cost  \\
\hline
\hline
10 & 0.3021 &  0.2698 \\
\hline
25 & 0.2743 &  0.2658 \\
\hline
50 & 0.2722 &  0.2656 \\
\hline
60 & 0.2730 &  0.2650 \\
\hline
80 & 0.2688 &  0.2657 \\
\hline
100 & 0.2673 &  0.2649 \\
\hline
\end{tabular}
\end{center}
\end{table}

\begin{table}
  \caption{Sensitivity of BBO to emigration rate function}
\label{table57}
\begin{center}
\begin{tabular}{ |c|c|c| } 
\hline
 & Linear  & Sinusoidal    \\
\hline
\hline
%\multirow{3}{4em}{Multiple row} & cell2 & cell3 \\ 
%& cell5 & cell6 \\ 
%& cell8 & cell9 \\ 
%\hline
Avg BBO Cost & 0.2696 &  0.3037 \\
\hline
Std Dev BBO Cost & 0.0014 &   0.0109\\
\hline
\end{tabular}
\end{center}
\end{table}

\begin{table}
  \caption{Sensitivity of PSO to $c_1$ and $c_2$}
\label{table66}
\begin{center}
\begin{tabular}{ |c|c|c|c|c|c| } 
\hline
 $c_1$ & 3  & 2 & 2 & 2 & 1  \\
 \hline
 $c_2$ & 2  & 3 & 2 & 1 & 2  \\
\hline
\hline
%\multirow{3}{4em}{Multiple row} & cell2 & cell3 \\ 
%& cell5 & cell6 \\ 
%& cell8 & cell9 \\ 
%\hline
Avg PSO Cost &  0.2657&  0.2653 &  0.2656& 0.2663 & 0.2655  \\
\hline
Std Dev PSO Cost & 0.0006 &  0.0005 & 0.0009 & 0.0007 & 0.0006  \\
\hline
\end{tabular}
\end{center}
\end{table}

\begin{table}
  \caption{Sensitivity of PSO to $w$}
\label{table45}
\begin{center}
\begin{tabular}{ |c|c|c|c|c| } 
\hline
 $w$ & 0.5  &  0.2 & -0.2 & -0.7   \\
\hline
\hline
%\multirow{3}{4em}{Multiple row} & cell2 & cell3 \\ 
%& cell5 & cell6 \\ 
%& cell8 & cell9 \\ 
%\hline
Avg PSO Cost & 0.2650 & 0.2645 & 0.2645 & 0.2646   \\
\hline
Std Dev Cost & 0.0003 & 0.0004 & 0.0002 &  0.0007  \\

\hline
\end{tabular}
\end{center}
\end{table}

\subsection{Performance metrics for comparing VE algorithms with NS algorithms}
Two performance criteria that are used to evaluate the performance of our Pareto-based optimization algorithms are normalized hypervolume and relative coverage~\cite{simon2013evolutionary}.\\
\textbf{Normalized Hypervolume: } This criteria is defined as
 $$S(\hat{p}_f)=\sum_{j=1}^{M} \prod_{i=1}^{k} f_i(x_j)/M$$ 
 where $M$ is the number of points in the Pareto front, $f$ is the $k$\_dimensional objective function of the optimization problem, and a smaller value of $S(\hat{p}_f)$ indicates better MOO performance.
 %, $\int_{0}^{Sim_{time}} |J|^2\ dt$, Max(Mag(J))

     \textbf{Relative Coverage:}
  Another way to compare Pareto front approximations is by computing the
number of individuals in one approximation that are weakly dominated by at least
one individual in the other approximation. The relative coverage for two Pareto fronts is computed as follows.
 \begin{equation}
 \label{eq:00}
\begin{split}
 & C(\hat{p}_{f}(1),\hat{p}_f(2))=\\
 & |a_2 \in \hat{p}_f(2); \exists a_1 \in \hat{p}_{f}(1): a_1 \geq a_2| 
  /|\hat{p}_f(2)|
  \end{split}
\end{equation}
Equation \ref{eq:00} defines the coverage
of $\hat{p}_{f}(1)$ relative to $\hat{p}_{f}(2)$ as the number of individuals in $\hat{p}_{f}(2)$ that are
weakly dominated by at least one individual in $\hat{p}_{f}(1)$.

\subsection{Comparison of VE algorithms with NS algorithms}
We  used  the  following  parameters  for NSPSO and VEPSO:
% To save space, I make it in one line

\begin{equation}
\label{eq:12}
\begin{split}
  G_s=7,\; \alpha=0.1,\; \beta=2,\; \gamma =2,\; \mu=0.1
  \end{split}
\end{equation}
where $G_s$, $\alpha$, $\beta$, $\gamma$ and $\mu$ are number of grids per dimension, inflation rate, leader selection pressure, deletion selection pressure, and mutation rate. Four different cost functions were uses, as introduced in Equation~\ref{eq:q8}.

Table~\ref{tabel417} shows the comparison of relative coverage between the multi-objective optimization algorithms. For instance, the entry 3/45 for the relative coverage of NSBBO to VEBBO means that 3 points out of 45 total NSBBO Pareto front points are weakly dominated by at least one point in the VEBBO Pareto front. Based on this table, NSBBO has better results in terms of smaller relative coverage, which means it has better a Pareto front set that is less dominated by the other Pareto front points.

    \begin{table}
\caption{Comparison of MOBBO with MOPSO in terms of relative coverage}

\label{tabel417}
\begin{center}
\begin{tabular}{ |c|c|c|c|c|c| } 
\hline
 & VEBBO & NSBBO & VEPSO & NSPSO   \\
\hline
\hline
%\multirow{3}{4em}{Multiple row} & cell2 & cell3 \\ 
%& cell5 & cell6 \\ 
%& cell8 & cell9 \\ 
%\hline
VEBBO & - & 3/45 & 7/20 & 0/20 \\
\hline

NSBBO & 15/47 &-& 5/20 & 0/20 \\
\hline

VEPSO & 1/47 & 0/45 &- & 6/20 \\
\hline
NSPSO & 21/47 & 0/45 & 0/20 &- \\
\hline
Tot. \% of Dom. Pts. & 79\% & 7\% & 35\% & 30\% \\
\hline
\end{tabular}
\end{center}
\end{table}

Table~\ref{tabel110} shows the simulation results relative to normalized hypervolume. It shows NSBBO works better in terms of smaller normalized hypervolume. 

\begin{table}
\caption{Comparison of MOBBO and MOPSO in terms of normalized hypervolume}
\label{tabel110}
\begin{center}
\begin{tabular}{ |c|c|c|c|c| } 
\hline
 & Normalized hypervolume & No. of Pareto Points \\
\hline
\hline
%\multirow{3}{4em}{Multiple row} & cell2 & cell3 \\ 
%& cell5 & cell6 \\ 
%& cell8 & cell9 \\ 
%\hline
VEBBO & 0.0024&47\\
\hline

\textbf{NSBBO} & 0.0016&45\\
\hline

VEPSO &0.0152 &20\\
\hline
NSPSO &0.0056 &20\\
\hline
\end{tabular}
\end{center}
\end{table}

\section{CONCLUSIONS}
In this paper, we chose BBO and PSO as two algorithms  that are representative of a typical evolutionary and swarm algorithm, respectively, and we extended them to multi-objective optimization to tune the parameters of the PD controller of a drone.
The multi-objective function is defined based on the tracking errors of the four states of the system, which in turn help reduce settling time, overshoot, rise time and steady state error.   VEBBO, NSBBO, VEPSO, NSPSO and aggregation methods were applied to the dynamic model of the drone.
Two evaluation criteria, normalized hypervolume and relative coverage, were used to compare the performance of the multi-objective optimization methods.
 The results showed improved results compared to conventional PD control. Also, we  investigated  the robustness,   sensitivity   and   repeatability   of the  aggregation method. To speed up operation, we used the MATLAB Parallel Computing Toolbox to parallelize the aggregation method.

For future work, we plan to conduct flight tests with an open source drone to test the controller with the optimized parameters we found with MOBBO and MOPSO. Also, we will investigate different weights for the aggregation method and we will investigate the effect of trajectory disturbances and noise in the PD parameters tuning process.
%In order to that, we need to find a method to choose one solution from the Pareto front set. 
%Also, we aim to implement a real-time framework to tune the parameters of the controller on-line and in real-time. 
%Then by using \textit{MATLAB Parallel Computing Toolbox}, we reduce the time of computing the BBO and PSO best cost

%%%%%%%%%%%%%%%%%%%%%%%%%%%%%%%%%%%%%%%%%%%%%%%%%%%%%%%%%%%%%%%%%%%%%%%%%%%%%%%%

%%%%%%%%%%%%%%%%%%%%%%%%%%%%%%%%%%%%%%%%%%%%%%%%%%%%%%%%%%%%%%%%%%%%%%%%%%%%%%%%

%%%%%%%%%%%%%%%%%%%%%%%%%%%%%%%%%%%%%%%%%%%%%%%%%%%%%%%%%%%%%%%%%%%%%%%%%%%%%%%%

%%%%%%%%%%%%%%%%%%%%%%%%%%%%%%%%%%%%%%%%%%%%%%%%%%%%%%%%%%%%%%%%%%%%%%%%%%%%%%%%

\bibliographystyle{IEEEtran} % Plain referencing style
\bibliography{azin} % Use the example bibliography file sample.bib
\end{document}